# Enhancing the evaluation of pathogen transmission risk in a hospital by merging hand-hygiene compliance and contact data: a proof-of-concept study

Rossana Mastrandrea[1], Alberto Soto-Aladro[2,3], Philippe Brouqui[2,3], Alain Barrat[1,4]

[1]Aix Marseille Université, Université de Toulon, CNRS, CPT, UMR 7332, 13288 Marseille, France
[2]Aix Marseille University, IHU Méditerranée Infection, URMITE, UM63, CNRS 7278, IRD 198, Inserm 1095, Marseille, France
[3]Infectious Disease Unit CHU Nord, Institut Hospitalo-Universitaire Méditerranée Infection, Marseille, France
[4]Data Science Laboratory, ISI Foundation, Turin, Italy

**Corresponding author**
Alain Barrat, CPT, Campus de Luminy Case 907, 13288 Marseille cedex 9, France
Tel: + 33 4 91 26 95 40 ; Fax: + 33 91 26 95 53; email: alain.barrat@cpt.univ-mrs.fr

**email addresses:**
R. Mastrandrea: rossmastrandrea@gmail.com
A. Soto-Aladro: alberto.soto-aladro@ap-hm.fr
Ph. Brouqui: philippe.brouqui@univ-amu.fr

**Abstract**

BACKGROUND. Hand-hygiene compliance and contacts of health-care workers largely determine the potential paths of pathogen transmission in hospital wards. We explored how the combination of data collected by two automated infrastructures based on wearable sensors and recording (i) use of hydro-alcoholic solution and (ii) contacts of health-care workers provide an enhanced view of the risk of transmission events in the ward.

METHODS. We perform a proof-of-concept observational study. Detailed data on contact patterns and hand-hygiene compliance of health-care workers were collected by wearable sensors over 12 days in an infectious disease unit of a hospital in Marseilles, France.

RESULTS. 10837 contact events among 10 doctors, 4 nurses, 4 nurses' aids and 4 housekeeping staff were recorded during the study. Most contacts took place among medical doctors. Aggregate contact durations were highly heterogeneous and the resulting contact network was highly structured. 510 visits of health-care workers to patients' rooms were recorded, with a low rate of hand-hygiene compliance. Both data sets were used to construct histories and statistics of contacts informed by the use of hydro-alcoholic solution, or lack thereof, of the involved health-care workers.

CONCLUSIONS. Hand-hygiene compliance data strongly enrich the information concerning contacts among health-care workers, by assigning a 'safe' or 'at-risk' value to each contact. The global contact network can thus be divided into 'at-risk' and 'safe' contact networks. The combined data could be of high relevance for outbreak investigation and to inform data-driven models of nosocomial disease spread.

BACKGROUND

Hospital Acquired Infections (HAI) represent a major risk for patient safety and health, leading to prolonged stay and increased mortality[1,2,3], and has high economic costs. The dynamics of pathogen transmission in hospital wards involves at least two components linked to the behavior of healthcare workers (HCWs): compliance to hand-hygiene protocols[4,5] and contacts between HCWs that might lead to transmission events[6,7]. In order to better understand such dynamics and ultimately develop prevention and containment measures, recent studies have concerned each of these two components.

On the one hand, detailed investigations of hand-hygiene compliance (HHC) of HCWs have been performed using techniques ranging from direct observation, still considered as a gold standard but costly, to sensors monitoring the use of hydro-alcoholic solution (HAS) by HCWs[8,9,10].

On the other hand, the recent development of unobstrusive wearable sensors has also made possible the collection of data describing the contact patterns of individuals with high spatial and temporal resolutions in various contexts[11,12,13,14], including hospital wards[15,16,17,18,19]. Moreover, the combination of contact data with microbiological data has been shown to provide evidence for the occurrence of transmission events of pathogens upon close proximity of individuals[18,19].

Contact data and hand-hygiene compliance data have however, to our knowledge, not been studied in combination, although numerical studies using agent-based models of hospital wards have clearly highlighted that both aspects play an important role in the potential spread of HAI outbreaks[6,7]. Here, we perform a first step in this direction through a proof-of-concept study in which we explore the merging of data coming from two data-collection infrastructures based on wearable sensors: MediHandTrace (MHT)[9], which monitors the use of HAS by HCWs, and the SocioPatterns infrastructure, which records the close face-to-face proximity of individuals. We show how combining such data provides an enhanced view of the risk of transmission events in the ward: indeed, the temporal network of contacts between HCWs and of visits of HCWs to patients' rooms results enriched by the information about hand-hygiene compliance of the individuals involved in each contact event. Such combined data could be of high relevance for outbreak investigation and to inform data-driven models of nosocomial disease spread.

METHODS

**Setting and Data Collection**

Data were collected from Monday, July 7, 2014 to Friday, July 18, 2014 in the Infectious Disease Unit of the Hôpital Nord in Marseilles, France. The study involved 7 rooms with patients and 22 healthcare workers (HCWs), categorized into 4 classes according to their role in the ward: 10 medical doctors (MED), 4 nurses (IDE), 4 nurses' aids (AS) and 4 housekeeping staff (ASH). For each day, each room was charaterized by a specific type of isolation requirement: "Standard" (no specific isolation procedure), "Contact", "Respiratory" and "Clostridium". For rooms in the "Clostridium" category, strong hygiene measures are required: HCWs are to wear gloves and special clothes before entering the room, downing gloves and clothes in the room before exit. Here, hand-hygiene compliance is considered as particularly important, especially after contact with the patient after removing gloves.

**Contact Data**

We recorded close proximity face-to-face interactions ("contacts") among HCWs in the ward, using the proximity-sensing platform developed by the SocioPatterns collaboration[11] and described in detail in [12,16].

Participating individuals were asked to wear unobstrusive badges on their chests, equipped with radio-frequency identification sensors that exchange ultra-low power radio data packets when close enough (1-1.5m). The system was tuned so that it detected and recorded close-range encounters during which a communicable disease infection could be transmitted, for example, by cough, sneeze or hand contact. The temporal resolution was set to 20s, and the information on face-to-face proximity events detected by the wearable sensors was relayed to radio receivers installed throughout the hospital ward. A sensor was also placed in the toilet just above the washbasin and close to the HAS dispenser. Contacts registered between this sensor and wearable sensors of HCWs were interpreted as a signal of hand-hygiene compliance.

**MediHandTrace (MHT) Data**

Seven patient rooms were equipped with antennas able to read sensors inserted in the shoes of 11 of the 22 HCWs (4 MED, 2 AS, 3 ASH and 2 IDE). The antennas were located around the HAS dispensers (one outside and one inside the room), at the room door and around the patient's bed[9]. Information about the presence in a room, the use of HAS, the vicinity of a patient of each participating HCW was transmitted in real-time to a centralized server.

**Data Analysis**

We first analyzed each data set separately. We studied the contacts between HCWs by focusing on contact matrices giving the number and durations of contacts between categories of HCWs and on the temporal evolution of these quantities during the day. We moreover built the daily and global networks of contacts between HCWs: in such networks, nodes represent HCWs and a link is drawn between two nodes if the corresponding individuals have met at least once in the period considered; each link is weighted by the total duration of the contacts between the concerned HCWs during the aggregation period.

We analyzed the MHT data by measuring, for each HCW category, the average number and duration of visits in patient rooms, as a function of the hour of the day. We moreover considered the (daily and global) bipartite network of visits between HCWs and rooms: nodes represent either HCWs or rooms, and a link is drawn between a HCW-node and a room-node if the corresponding HCW visited the room at least once; each link is weighted by the total duration of the visit(s) and can moreover be characterized by the HHC of the HCW during the visit. We also investigated the overall and individual HHC of HCWs.

Merging the two data sets yields a temporally resolved contact network between HCWs in which each contact is enriched by the information about the HHC of the involved HCWs before the contact. We considered specific examples of HCWs contact histories and explored how HCWs' contacts can be divided into 'at-risk' and 'safe' categories. We built the corresponding daily and global aggregated contact networks.

**Ethics and Privacy**

An oral and written consent to participate in the study was obtained from the HCWs. In order to ensure anonymity of the data, a random number was attributed to each participant when entered in the database. This study was declared to the French Commission on Individual Data Protection and Public Liberties (CNIL).

## RESULTS AND DISCUSSION

### Contact Data

A total of 10837 contact events was recorded during the study, with a cumulative duration equal to 453960s (~7566 min or 126h). The durations of single contacts and the aggregated durations of contacts between HCWs were highly heterogeneous (see Supplementary Material): most contacts had very short durations, but some were very long, with no characteristic interaction time-scale. Most contact activity occurred within the MED category on the one hand and between the AS and ASH categories on the other hand (Fig. 1a). We show in the Supplementary Material that daily contact matrices exhibit the same structure as the global one. Contacts occurred mostly between 10:00 and 14:00 for MEDs and IDEs, while AS-ASH contacts spread more evenly along the course of the day (Fig. 1b).

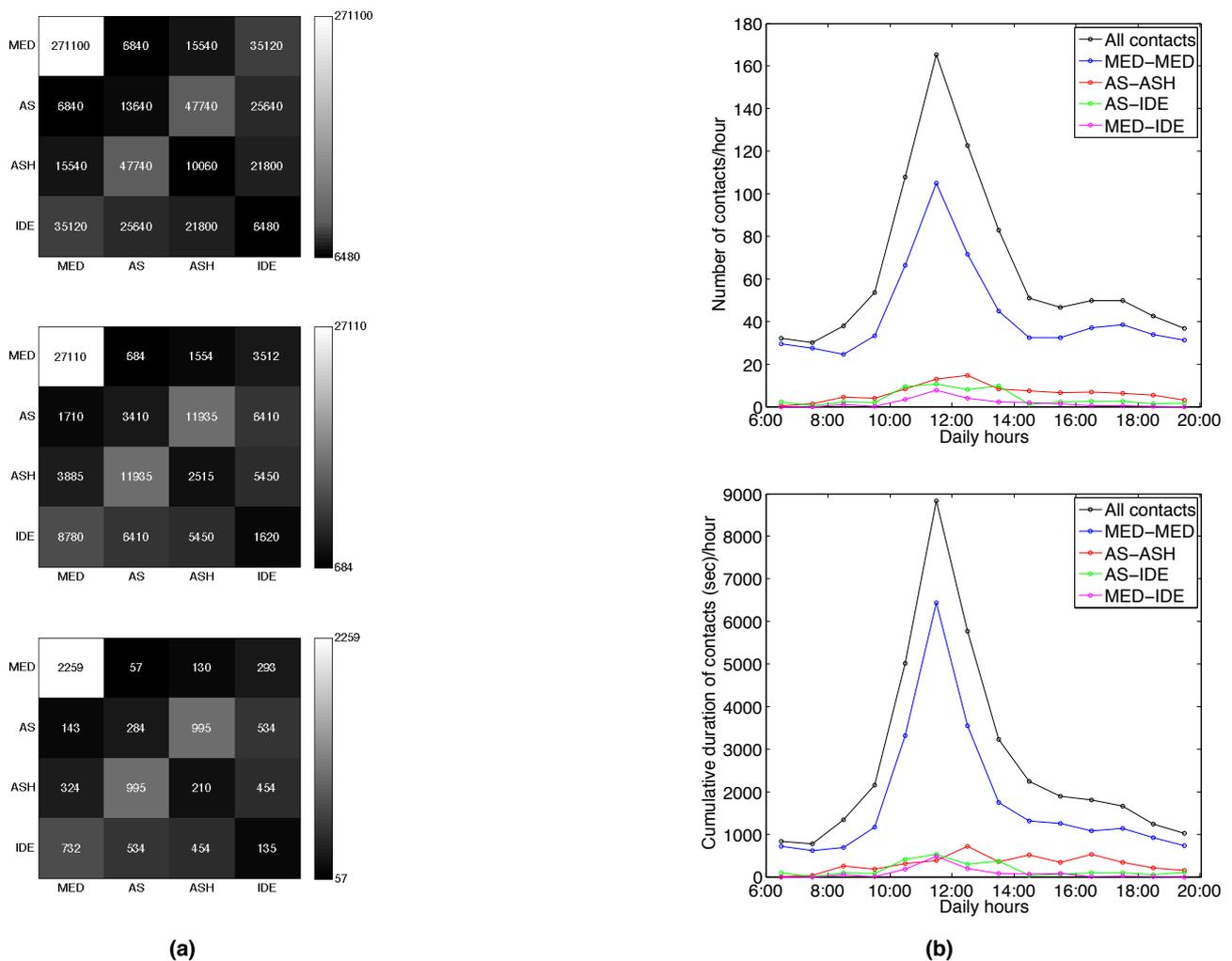

**Figure 1. Contacts between health-care workers.** (a) Contact matrices: the top matrix gives at row X and column Y the aggregate duration of contacts between HCWs of category X and HCWs of category Y; in the second matrix, each (X,Y) entry is normalized by the number of HCWs in category X, giving thus the average time spent by a HCW of category X with all HCWs of category Y; in the bottom matrix, each entry is moreover normalized by the duration (in days) of the data collection. (b) Number (top) and aggregate duration (bottom) of contacts as a function of the hour of the day, averaged over all the days of data collection. Data are shown for all contacts and for some specific categories. Abbreviations: MED, medical doctors; IDE, nurses; AS, nurses' aids; ASH, housekeeping staff.

More detailed insight into the contact patterns is provided by the contact networks of HCWs: the global aggregated network is shown in Fig. 2a) using a circular layout with HCWs grouped by category (daily networks are shown in the Supplementary Material). Overall, MEDs form a well-connected group connected to the rest of the network mostly through contacts with a couple of IDEs. Figure 2b) provides network statistics for each HCW, namely, degree (i.e., the number of other HCWs with whom a given HCW has had contacts) and strength (i.e., cumulative contact time of each HCW). While the degrees do not exhibit very strong variations (as in other studies[12,16,17]), different HCWs can have very different cumulative contact times. In particular, only 5 (resp. 10) HCWs out of 22 account for 44% (resp. 72%) of the total contact time of all HCWs. Figure 2c) finally shows the possibility to focus on the contact history of a specific HCW and to capture at a glance how distinct contact patterns occured at specific periods: for instance, HCW 113 (ASH) met repeatedly HCW 127 (AS) in different days, and met HCW 67 (MED) for a long time but just on one day.

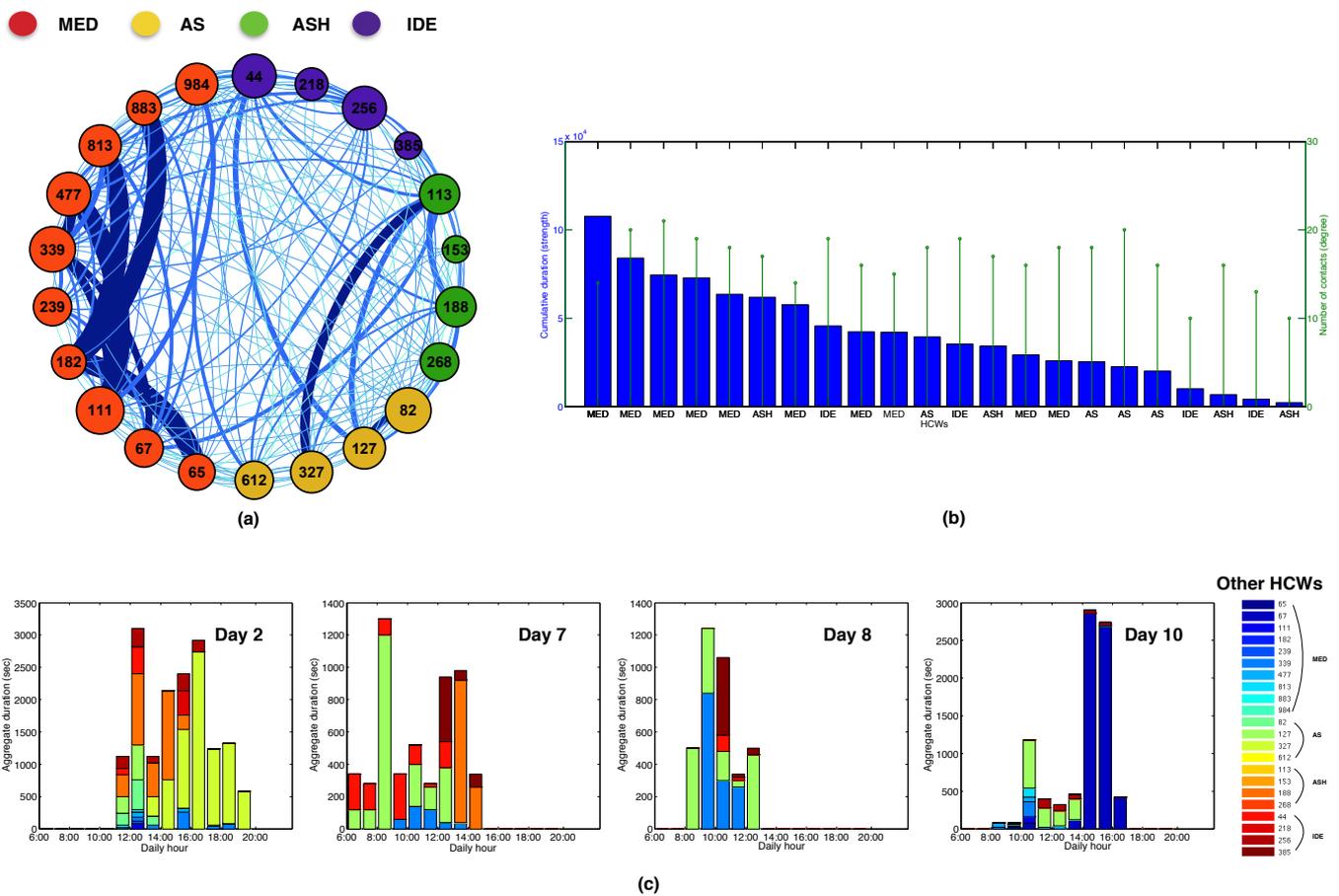

**Figure 2. Contact network of health-care workers.** (a) Visualization of the contact network of HCWs, aggregated over the whole data collection. Each node represents a HCW (the number stands for the RFID number, the color for the category as indicated in the color code). A link between two nodes exists if at least one contact between the corresponding HCWs was detected, and the link thickness gives the aggregate duration of contacts. The size of each node is given by its degree (number of links). (b) Cumulative duration (blue bars) of the contacts of each HCW, and number (green) of distinct other HCWs met by each HCW, during the whole data collection. These quantities correspond respectively to the strength and the degree of each node in the network of Figure 2(a). (c) Contact patterns of a specific HCW along several days. Each bar gives, for each hour of the day, the cumulative duration of contacts between HCW 113 (ASH) and other HCWs identified each by a specific color. Abbreviations: MED, medical doctors; IDE, nurses; AS, nurses' aids; ASH, housekeeping staff.

## Visits to patients and Hand-Hygiene Compliance

During the data collection, 510 visits of HCWs to patients were registered with an average duration of 247s (~ 4 min). Figure 3a) shows that most visits occured in the first part of the day, with a peak before noon. Most visits were performed by AS and ASH, while MEDs performed only few visits, always around 1:00PM. Figure 3b) displays the bipartite network involving HCWs and rooms, aggregated over all the study (daily networks are shown in the Supplementary Material): each link indicates that a HCW has visited a room at least once, and its width indicates the total (cumulated) durations of the visits. The network is dense, with 63 links out of the 77 possible (82%). The degree of HCW nodes (number of different rooms visited) varies between 2 and 7, while the degree of room nodes (number of distinct HCWs who have visited the room) goes from 7 to 11. The figure highlights the weak participation of MEDs to visits and, on the opposite, the large cumulated time spent by some HCWs in some rooms.

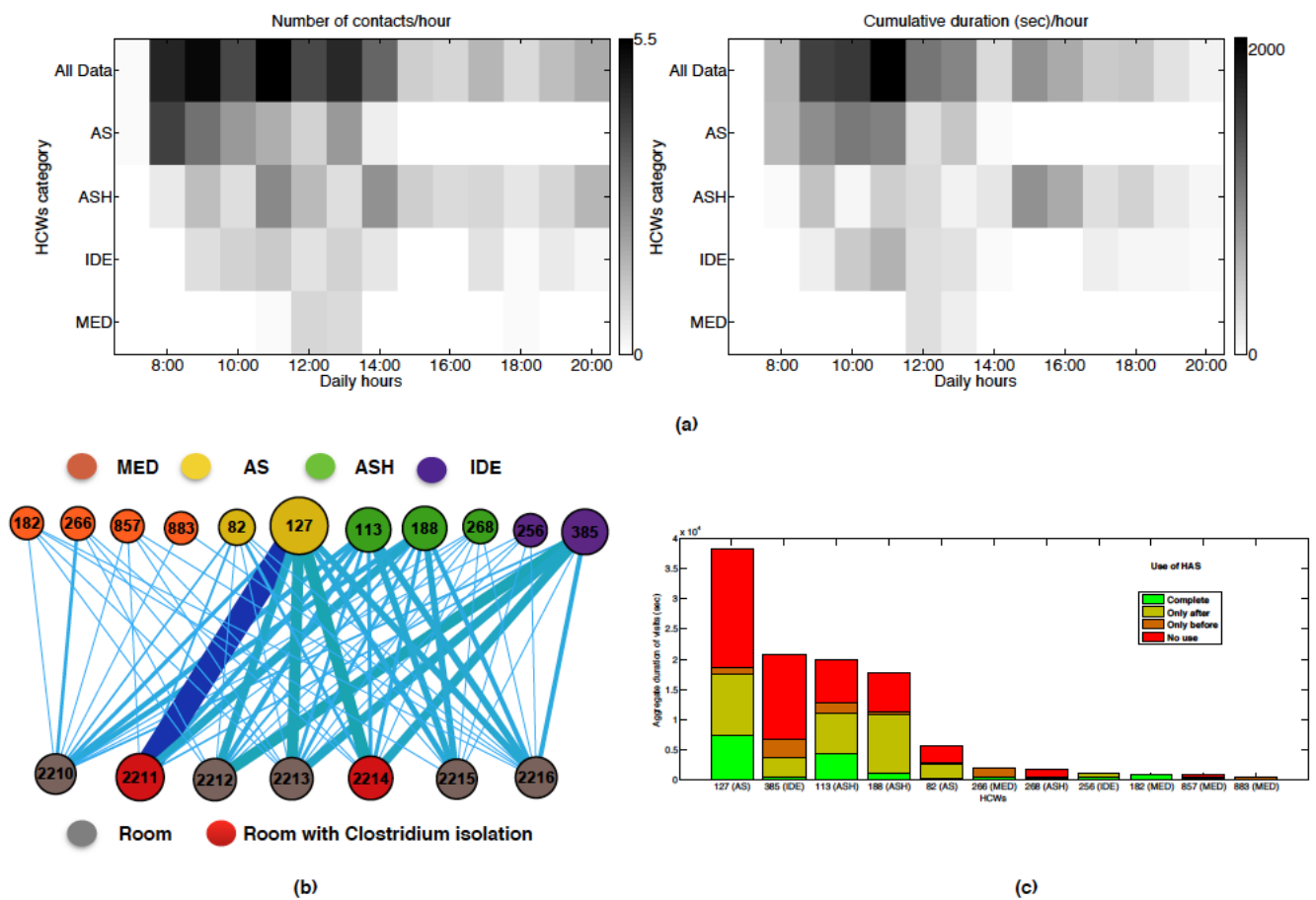

**Figure 3. MediHandTrace data: visits of health-care workers to patient rooms and hand-hygiene compliance.** (a) Number (left) and aggregate durations (right) of visits of HCWs to patient rooms per hour of the day, averaged over all days of the data collection, for all HCWs and for each HCW category. (b) Bipartite network of visits of HCWs to patient rooms. Each node represents either a HCW (top row) or a room (bottom row). The size of a node depends on its strength (aggregate duration of visits). The color of each HCW-node indicates its category, and red room-nodes indicate that the corresponding room was in "Clostridium" isolation during the whole data collection period. A link is drawn between a HCW and a room if the HCW visited the room at least once, and the link thickness depends on the aggregate duration of the visits. (c) Each bar gives the aggregated duration of all visits performed by a HCW. The total duration is split in different colors depending on the correct or incorrect use of HAS during the visits (correct use, use only before or after the contact with the patient, or no use at all). Abbreviations: MED, medical doctors; IDE, nurses; AS, nurses' aids; ASH, housekeeping staff; HAS, hydroalcoholic solution.

For each visit, the MHT infrastructure gave moreover information on the occurrence of an event of proximity with the patient ("contact") and on the hand-hygiene compliance of the HCW. Figure 3c) provides statistics on the use of HAS of each HCW before and/or after contact with a patient (further statistics can be found in the SI). Consistently with previous studies[8,9,21,22], completely correct use of HAS was rare (13%), while many visits (49%) occurred with no HAS use at all. The remaining visits were characterized by a partial use, mostly after contact with the patient: 8% (resp. 30%) with HAS use before (resp. after) the contact. We did not observe any clear trend of the hand-hygiene compliance as a function of the number or durations of visits, nor depending on the type of isolation of the room (data in Supplementary Material). Overall, hand-hygiene compliance was scarse on average for all HCW categories, but important differences were observed between HCWs (see Supplementary Material).

**Merging the two datasets**

We merged the data sets describing HCWs' visits to patients' rooms, their hand-hygiene compliance and their contacts, obtaining a detailed time-resolved picture of daily paths and contacts of HCWs with associated risks of hand transmission of pathogens due to the lack of hand-hygiene. The merged data concerns 9 HCWs: 2 MEDs, 2 AS, 3 ASH, 2 IDE. Figure 4a) illustrates the richness of the combined data by showing visits, hand-hygiene and contacts of a specific HCW (113): contacts of 113 occurring between two room visits are shown together, with a bar height proportional to the contacts' durations, and the colored dot on top of the bar indicates if 113 used HAS or not after contact with a patient and before contact with the other HCWs. Information about the use of HAS by the HCW with whom 113 had a contact is also included.

We now focus on the contacts between two HCWs in which at least one of them participated to the MHT data collection, and such that the encounter occurred after a contact with a patient. This corresponds to 2593 contact events (24% of all contact events) of total duration 132420s (29% of the aggregated duration of all contacts) and concerning 92 of the 182 links between HCWs (51%). To each such contact, we can associate a risk of hand transmission of pathogens, depending on the HCW's hand-hygiene compliance (note that we do not take into account here the contacts occurring at the start of the HCWs shifts, before they have any contact with patients). Figure 4 reports the corresponding statistics at the level of pairs of HCWs (b) and of individual HCWs (c). Each bar gives the total duration of the measured contacts between a pair of HCWs (resp. of all contacts of each HCW), and is divided into durations of 'at-risk' and 'safe' contacts, according to the use of HAS by the HCWs before the contact. The average aggregate duration of 'safe' contacts was larger than the one of 'at-risk' contacts (286s vs. 190s), but large variations, with both short and long contacts, were observed in each case. We did not observe any significant correlation between total contact durations and hand-hygiene compliance of HCWs. Overall, 7 HCWs out of 9 had contacts with other HCWs after contact with a patient and lack of HAS use.

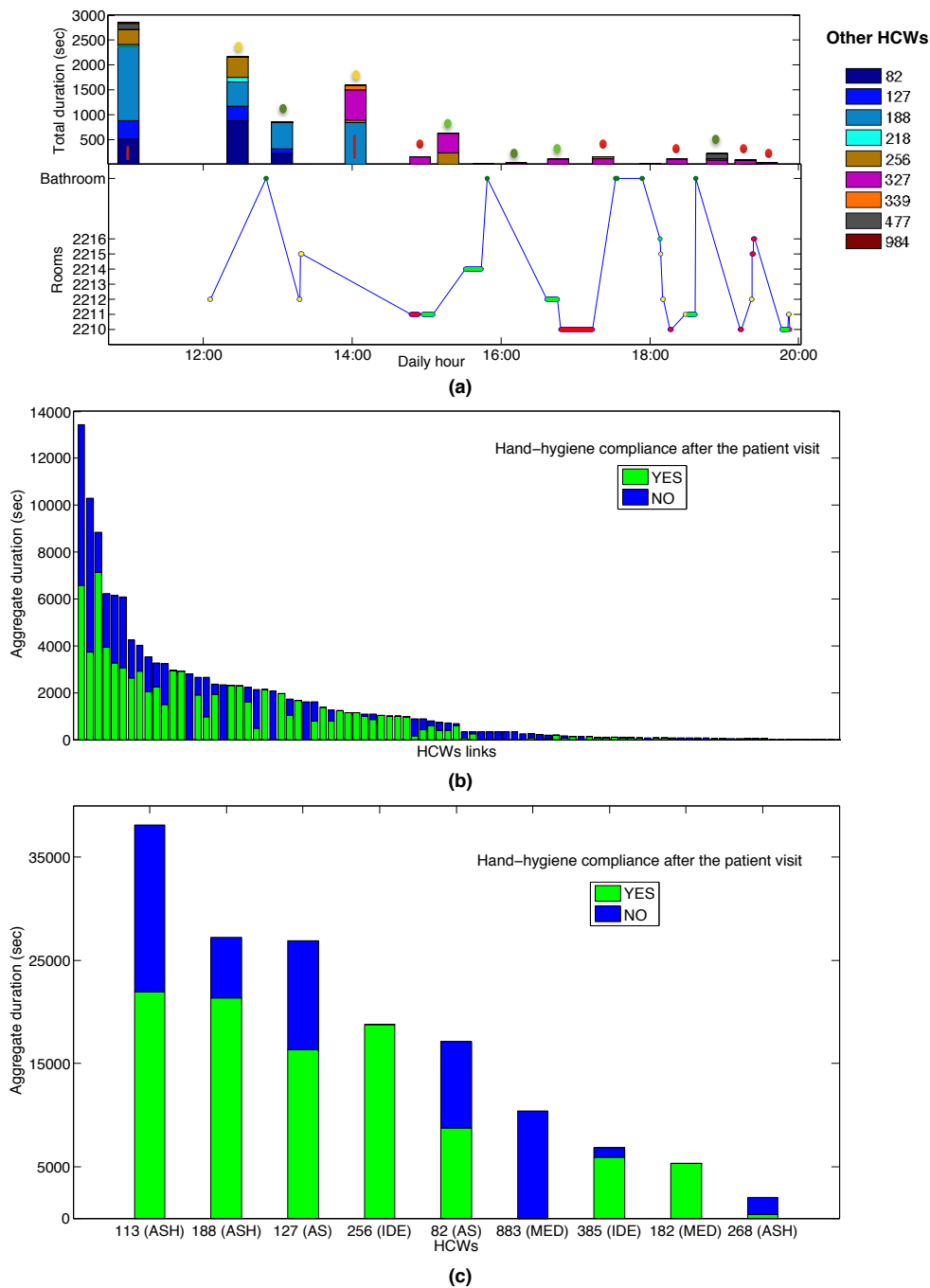

**Figure 4. HCW contacts and hand-hygiene compliance.** (a) Contact patterns of a single HCW (113, ASH) during a specific day, informed by hand-hygiene compliance data. Bottom: each dot represents a visit to a room, and its color depends on the use of HAS, or lack thereof, after a contact with the patient in the room (red: no use; light green: use; dark green: bathroom; yellow: no contact with the patient). Top: each bar gives the duration of a contact of HCW 113 with another HCW, identified by a specific color. The colored dot indicates if HCW 113 had used HAS in the previous visit (same color code as in the bottom plot). The red line in some colored bars indicate that the HCW in contact with HCW 113 had previously visited a patient room with successive lack of HAS use. (b) Bar chart giving the aggregate duration of the contacts of each pair of HCWs. Only encounters between HCWs of which at least one wore a MediHandTrace (MHT) tag monitoring hand-hygiene compliance, and had a contact with a patient before the encounter, are considered. Each duration is divided into two parts corresponding to different hand-hygiene compliance (green: correct use of HAS of the HCWs monitored by the MHT infrastructure before the contact; blue: no use of HAS after contact with the patient and before the contact for at least one of the two HCWs involved). (c) Bar chart giving, for each HCW, the aggregate duration of his/her contacts with other HCWs after contact with a patient. Each aggregate duration is divided into two parts corresponding to different hand-hygiene compliance (green: correct use of HAS after contact with the patient and before the contact with the other HCW; blue: no use of HAS). Abbreviations: MED, medical doctors; IDE, nurses; AS, nurses' aids; ASH, housekeeping staff; HAS, hydroalcoholic solution.

Figure 5 finally illustrates how networks of contacts between HCWs can be enriched by the HHC data and divided into two networks corresponding to 'at-risk' and 'safe' contacts. We show in Fig. 5 the global contact network and two daily networks, each one split according to this criterion. In each layout, nodes are placed in the same location, and the link thickness corresponds to the aggregated duration of the contacts between the linked nodes. 'At-risk' and 'safe' contact networks show different patterns, with some links present in both while others are observed only in one. In particular, only the 'at-risk' network should be considered in a hypothetical scenario of hand transmission of pathogens. Risk assessment analysis should thus focus on this network and its characteristics. We also emphasize that Fig. 5 gives only static representations of a temporal network, in which each link represents an aggregation of several contact events. The whole data set includes the temporality of contacts together with their risk status in terms of hand-hygiene of the involved HCWs.

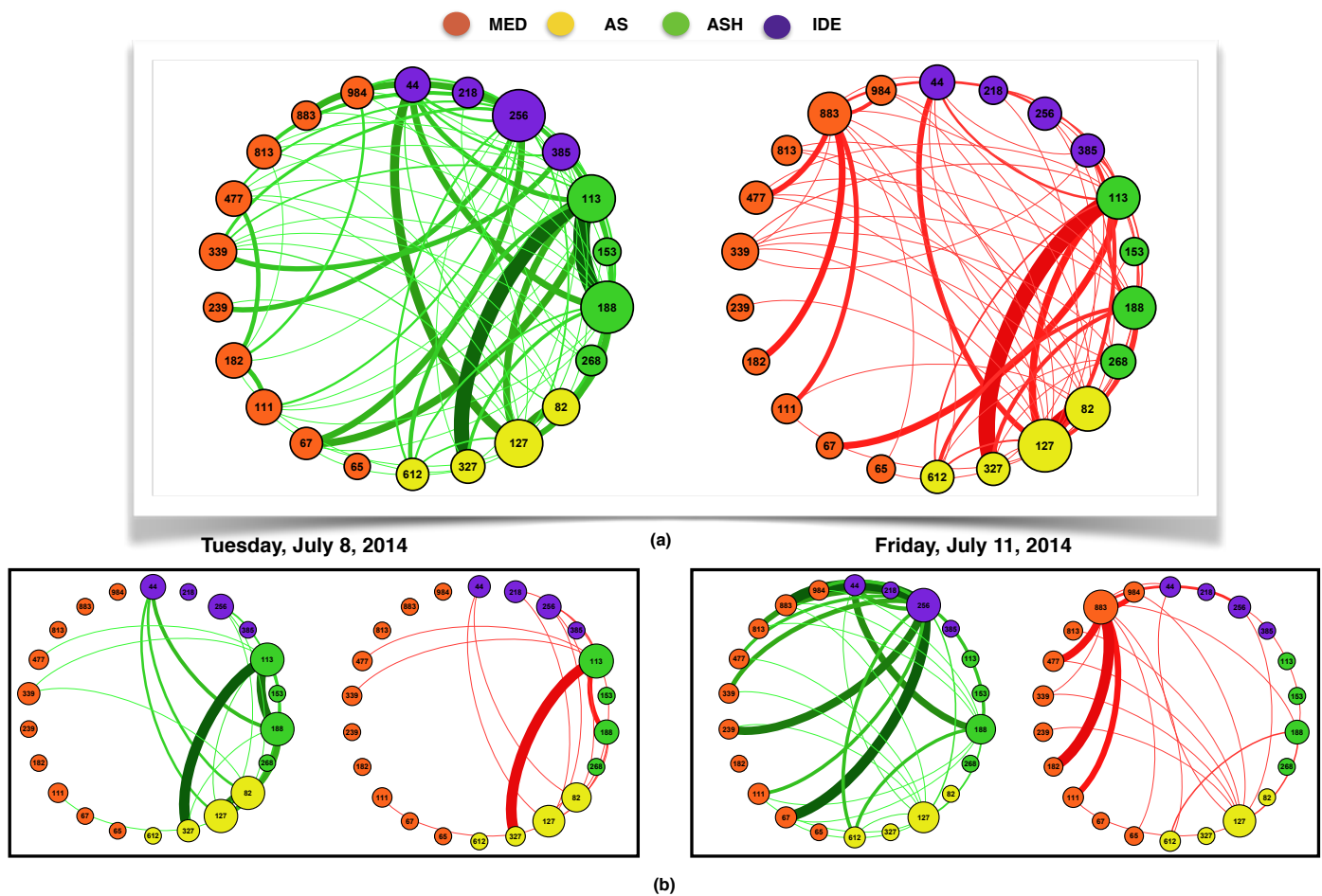

**Figure 5. Networks of HCWs contacts informed by the hand-hygiene compliance data.** In each network, similarly to Figure 2(a), nodes represent HCWs, with a different color according to each HCW's category, and a link between two nodes indicates that the corresponding HCWs have been in contact at least once. Here only contacts for which at least one of the HCWs was monitored by the MediHandTrace (MHT) infrastructure, and occurring after at least one of the HCWs involved had a contact with a patient, are taken into account. We create two contact networks by integrating the information about hand-hygiene compliance of HCWs: the network of green links corresponds to contacts in which HCWs monitored by MHT have used HAS before the contact; the network of red links corresponds to contacts occurring after at least one of the HCWs had a contact with a patient and did not use HAS afterwards. In each network, the links thickness depends on the aggregate duration of contacts. (a) Networks aggregated over the whole data collection; (b) Two examples of daily contact networks. Abbreviations: MED, medical doctors; IDE, nurses; AS, nurses' aids; ASH, housekeeping staff, HAS, hydroalcoholic solution.

DISCUSSION

The objective of this work was to show how merging data concerning contacts and hand-hygiene compliance of health-care workers in a hospital ward can provide an enriched picture of transmission risks in such contexts. Previous data collection and data-driven studies focused indeed typically on only one of these aspects. For each separate data set, we obtained results consistent with previous works. For instance, the duration of contacts was heterogeneous, with the coexistence of contacts of very different durations and the emergence of "supercontactors" who account for a large fraction of all observed contacts[12,14,16,17]. In terms of contagion risk, this implies that different contacts can correspond to very different probabilities of transmitting a pathogen. The contact network was moreover strongly structured as in other studies[12,14,16,17]. These results confirm the need to go beyond the contact matrix description that only accounts for averaged contact durations between different categories of workers[20] and to include information about contact network structure and heterogeneity of contacts duration in data-driven models of disease spreading or in outbreak investigations.

Differently from other studies, most contacts were observed among medical doctors and between nurses'aids and housekeeping staff, due to a specific organization of work in the ward with pairs of one AS and one ASH working together.

With regards to hand-hygiene compliance, results were in line with other studies[8,9,21,22], with only 13% of visits of HCWs to room patients including a correct use of HAS (both before and after contact with the patient). The widely spread lack of hand-hygiene compliance emphasizes the interest of merging the two data sets: on the one hand, the risk of pathogen transmission during a contact between HCWs can strongly depend on the occurrence of visits to patients and of the corresponding use of HAS (or lack thereof) after contact with a patient and before the HCW-HCW contact; on the other hand, it shows how lack of HAS use should not be considered as an isolated phenomenon but can have far-reaching consequences as pathogens transmitted from a patient to a HCW could then easily be transmitted through the HCW contact network and spread throughout the ward. It emphasizes how each contact with a patient and each hand-washing event are not isolated events but part of a complex network of events.

We have illustrated how simple visualizations of HCWs visits and contacts can be obtained by such data and could be of use in outbreak investigation. We have also shown how the network analysis of HCWs' contacts is enriched by taking into account hand-hygiene data. The network could be made even richer by combining the networks of HCWs' contacts and visits, and assigning a risk to each link according to the use of HAS. Note that in the present proof-of-concept study, we have focused on use of HAS after contact with a patient, which determines the risk of HCW-HCW contacts, while such an enriched network could also encode information on the use of HAS before the contact with a patient in the links between room-nodes and HCW-nodes, as illustrated in Fig. 6. This network thus includes information not only on the risk of pathogen transmission between HCWs but also from HCWs to patients and provides a picture of the possible paths of pathogen transmission in the ward that would be more complete than usual descriptions based only on contact patterns (due to the complexity of the representation, we show in Fig. 6 only the part of the network around a specific HCW.)

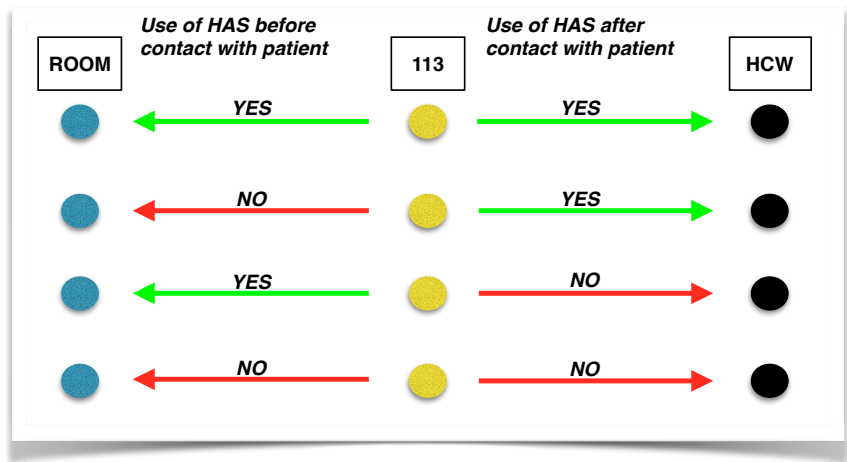

(a)

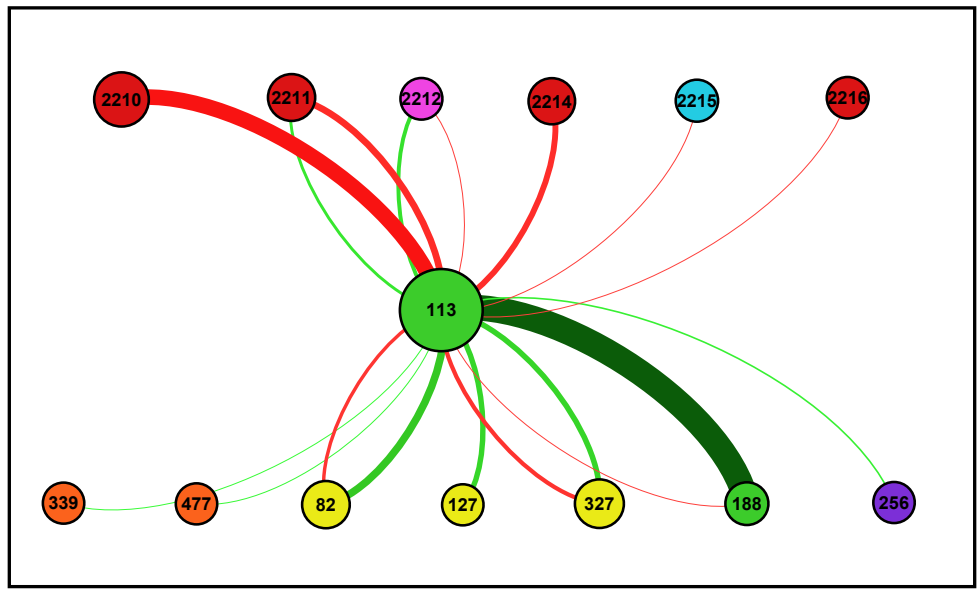

(b)

**Figure 6. Illustration of the complexity of the network including HCWs' contacts and rooms' visits.** (a) An example of the different types of contacts involving a specific healthcare worker after a visit to a patient. The hand-hygiene compliance can affect both HCWs and patients, with four possible scenarios according to the use of HAS before and after the visit, as illustrated in the figure: the first line corresponds to HCW 113 using HAS before contact with a patient (green arrow from 113 towards the room) and also after the contact with the patient and before the contact with the other HCW (green arrow towards the other HCW); the second line corresponds to HCW 113 having a contact with the patient in the room without previous use of HAS (hence a risk of transmission to the patient – red arrow) but using HAS after the contact with the patient and before contact with the other HCW. Lines 3 and 4 correspond to the scenarios of lines 1 and 2 respectively, but without use of HAS after the contact with the patient. (b) Network of all the contacts between a specific HCW (113), patients (indicated by their room number) and other HCWs during a specific day. A link between two nodes exists if the contact was observed at least once in the day; green and red links correspond to the use (or lack thereof) of HAS before the contact with the patient or between contact with patient and contact with HCW; the thickness of each link stands for the cumulative duration of the corresponding contacts. This figure integrates the information about at-risk and safe contacts of Figure 5 (a) with the information about the use of HAS before the visit to the patient (i.e., the risk associated to the encounters HCW-patient).

CONCLUSION

Hand-hygiene compliance data strongly enrich the information concerning contacts among health-care workers: indeed, each contact can be assigning a 'safe' or 'at-risk', yielding two distinct 'at-risk' and 'safe' contact networks. The resulting enriched temporal contact network could be of high relevance for outbreak investigation and to inform data-driven models of nosocomial disease spread.
Further work includes repeated data collections and analysis in order to test the robustness of the statistical patterns observed, as well as the design of data-driven models informed by the collected data to simulate the potential spread of pathogens and identify efficient containment measures in case of an outbreak. The results presented here will also be used to provide feedback to HCWs and attract their attention on the complex interplay of contacts and hand-washing events and how lack of hand-hygiene compliance can potentially lead to consequences involving the whole network of HCWs. As the question of how to increase hand-hygiene compliance among HCWs is still wide open[21], the effect of such feedback could then be measured in a new data collection campaign.

LIST OF ABBREVIATIONS

HCW, health-care worker; HHC, hand-hygiene compliance; HAS, hydro-alcoholic solution; MHT, MediHandTrace; MED, medical doctors; IDE, nurses; AS, nurses' aids; ASH, housekeeping staff.

COMPETING INTERESTS

The authors declare that they have no competing interests.

AUTHORS' CONTRIBUTIONS

AB and PB conceived and designed the study. AB, PB, AS, RM deployed the infrastructures and collected the data. RM analyzed the data, performed the statistical analysis, created the figures and wrote the first draft of the manuscript. AB, PB and RM wrote the final version of the manuscript. All authors read and approved the final manuscript.

ACKNOWLEDGMENTS
We thank the SocioPatterns collaboration for providing access to the SocioPatterns sensing platform that was used in collecting the contact data. We are grateful to the hospital staff who volunteered to participate in the data collection. This work has been carried out thanks to the support of the A*MIDEX project (n° ANR-11-IDEX-0001-02) funded by the « Investissements d'Avenir » French Government program, managed by the French National Research Agency (ANR). A.B. is partially supported by the French ANR project HarMS-flu (ANR-12-MONU-0018) and by the EU FET project Multiplex 317532. The authors have no conflict of interest. The funding bodies had no role in the design of the study nor in the data analysis and interpretation.

**Supplementary Material**

**Figure S1. Distribution of aggregate durations of contacts.** Distribution of the aggregated durations of contacts between HCWs, computed over the whole data collection.

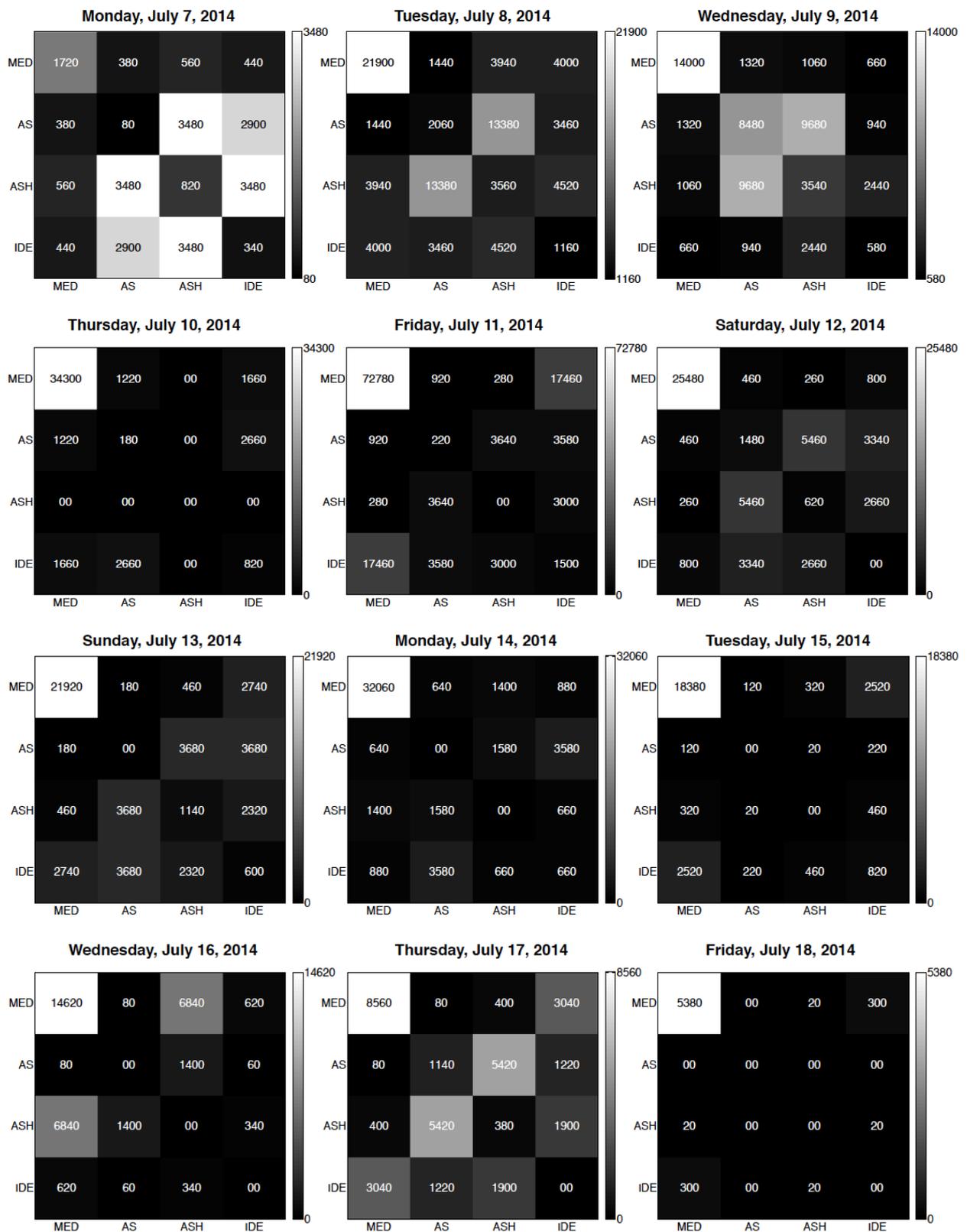

**Figure S2. Daily contact matrices.** Each matrix gives at row X and column Y the aggregate duration of contacts between HCWs of category X and HCWs of category Y during a specific day.
Abbreviations: MED, medical doctors; AS, nurses; ASH, nurses' aids; IDE, housekeeping staff.

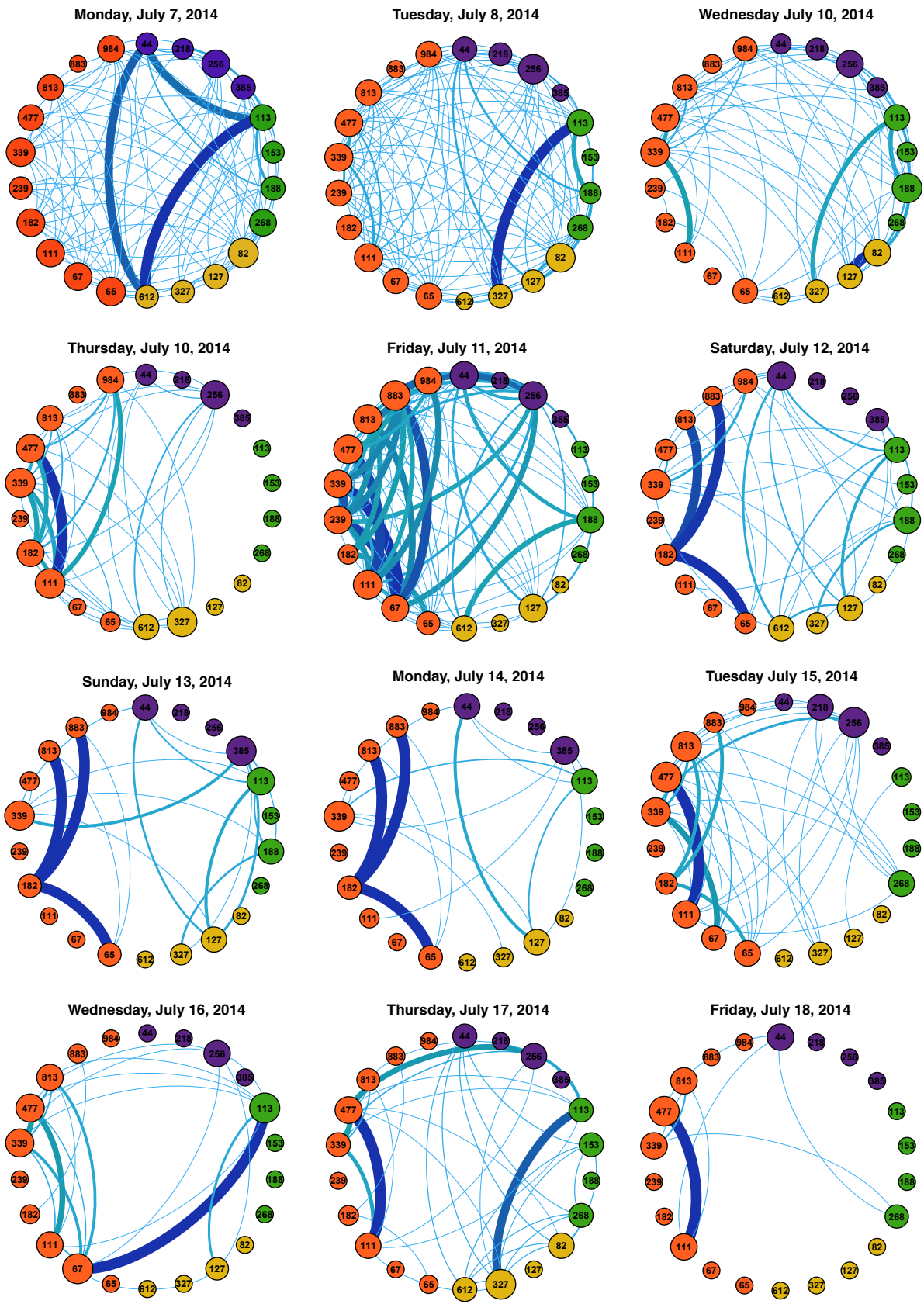

**Figure S3. Daily contact networks of health-care workers.** Visualization of the contact network of HCWs, aggregated at a daily scale. Each node represents a HCW (the number stands for the tag number, the color for the category as indicated in the color code). A link between two nodes exists if at least one contact between the corresponding HCWs was detected, and the link thickness gives the aggregate duration of contacts during the considered day. The size of each node is given by its degree (number of links). Abbreviations: MED, medical doctors; AS, nurses; ASH, nurses' aids; IDE, housekeeping staff.

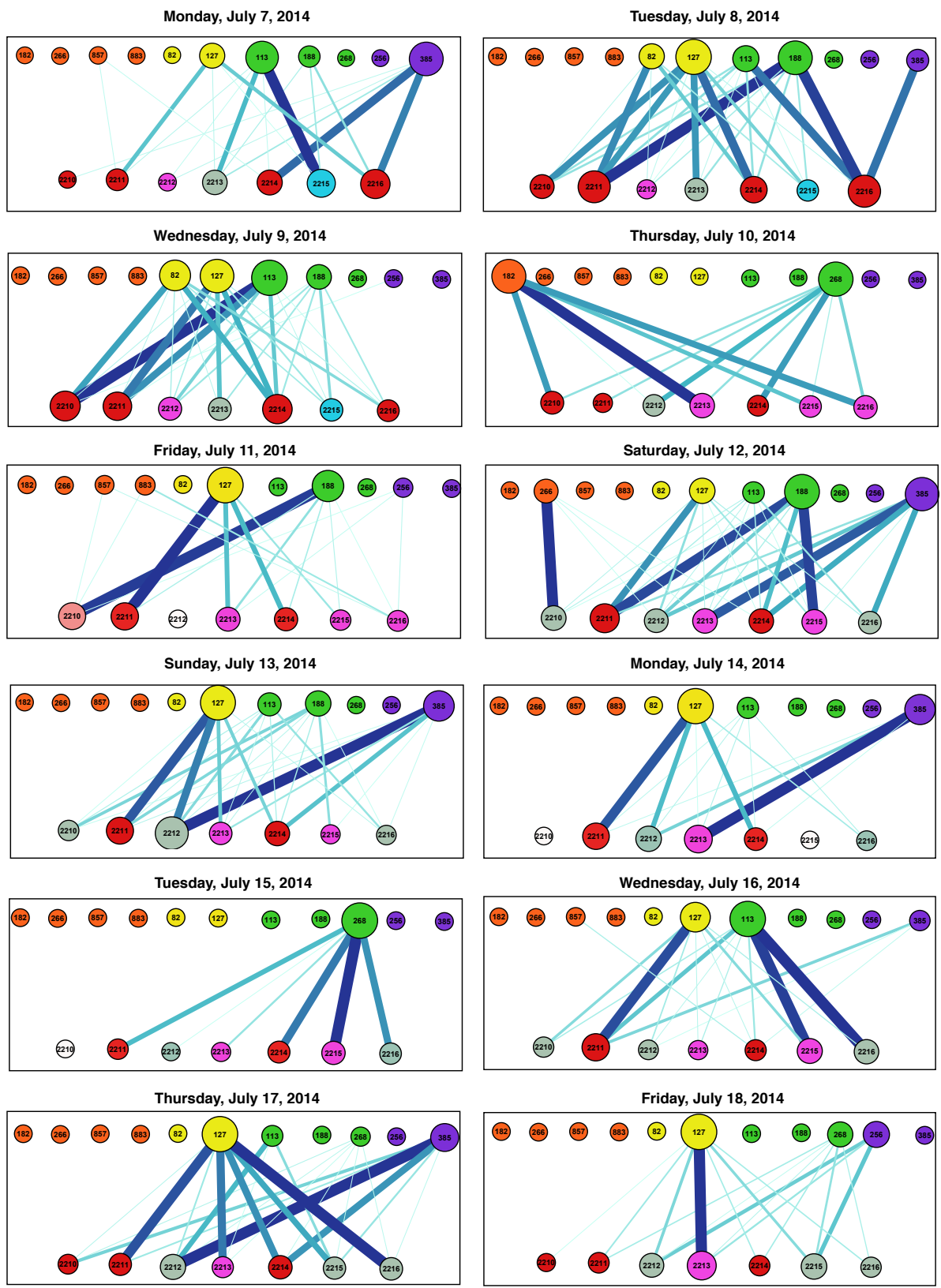

**Figure S4. Daily bipartite networks of contacts between health-care workers and patients.** Bipartite network of visits of HCWs to patient rooms. Each node represents either a HCW (top rows of each network) or a room (bottom rows). A link is drawn between a HCW and a room if the HCW visited the room at least once, and the link thickness depends on the aggregate duration of the visits. The size of a node depends on its strength (aggregate duration of visits). The color of each HCW node indicates its category, the color of each room node represents the type of isolation. Abbreviations: MED, medical doctors; AS, nurses; ASH, nurses' aids; IDE, housekeeping staff.

## (a)

| MHT data | |
|---|---|
| **Total visits** | 510 (321) |
| **Visits without contact with patient** | 189 (37%) |
| **Visits with contact and no use of HAS** | 156 (48.5%) (30.5%) |
| **Visits with contact and use of HAS only BEFORE** | 26 (8%) (5%) |
| **Visits with contact and use of HAS only AFTER** | 95 (29.5%) (19%) |
| **Visits with contact and complete use of HAS** | 44 (14%) (8.5%) |

(a)

## (b)

| | 82 (AS) | 113 (ASH) | 127 (AS) | 182 (MED) | 188 (ASH) | 256 (IDE) | 266 (MED) | 268 (ASH) | 385 (IDE) | 857 (MED) | 883 (MED) |
|---|---|---|---|---|---|---|---|---|---|---|---|
| Visits with contact and no use of HAS | 12 (50%) | 22 (37%) | 47 (47.5%) | 0 (0%) | 22 (42%) | 0 (0%) | 1 (50%) | 11 (79%) | 39 (76%) | 2 (50%) | 1 (50%) |
| Visits with contact and use of HAS only BEFORE | 4 (17%) | 7 (11.5%) | 6 (6.5%) | 0 (0%) | 2 (4%) | 0 (0%) | 1 (50%) | 1 (7%) | 3 (6%) | 1 (25%) | 1 (50%) |
| Visits with contact and use of HAS only AFTER | 7 (29%) | 19 (31.5%) | 31 (31%) | 0 (0%) | 23 (44%) | 6 (67%) | 0 (0%) | 2 (14%) | 7 (14%) | 0 (0%) | 0 (0%) |
| Visits with contact and complete use of HAS | 1 (4%) | 12 (20%) | 15 (15%) | 5 (100%) | 5 (10%) | 2 (23%) | 0 (0%) | 0 (0%) | 2 (4%) | 1 (25%) | 0 (0%) |
| Total Visits | 24 (100%) | 60 (100%) | 99 (100%) | 5 (100%) | 52 (100%) | 8 (100%) | 2 (100%) | 14 (100%) | 51 (100%) | 4 (100%) | 2 (100%) |
| Visits without contact | 13 | 31 | 66 | 0 | 36 | 1 | 3 | 12 | 28 | 2 | 0 |

(b)

**Figure S5. Hand-hygiene compliance of HCWs.** (a) Total number of visits to patients in the whole period of data collection, divided according to the use of hydro-alcoholic solution before or/and after the contact with the patient. The numbers in blue are computed when taking into account only the visits that included a contact with the patient. (b) Number of visits to patients divided according to the use of HAS for each HCW monitored by the MHT system. The percentages are computed with respect to columns and do not take into account the visits without contacts with the patient.
Abbreviations: MED, medical doctors; AS, nurses; ASH, nurses' aids; IDE, housekeeping staff.

| ISOLATION ──────── USE of HAS | STANDARD | CONTACT | RESPIRATORY | CLOSTRIDIUM | TOTAL |
|---|---|---|---|---|---|
| COMPLETE | 11 (13%) | 2 (14%) | 12 (14%) | 19 (14%) | 44 (13%) |
| ONLY BEFORE | 8 (10%) | 3 (21.5%) | 7 (8%) | 8 (6%) | 26 (8%) |
| ONLY AFTER | 22 (26%) | 3 (21.5%) | 22 (27%) | 48 (34%) | 95 (30%) |
| NO USE | 43 (51%) | 6 (43%) | 42 (51%) | 65 (46%) | 156 (49%) |
| TOTAL | 84 (100%) | 14 (100%) | 83 (100%) | 140 (100%) | 321 (100%) |

(a)

| ISOLATION ──────── USE of HAS | STANDARD | CONTACT | RESPIRATORY | CLOSTRIDIUM | TOTAL |
|---|---|---|---|---|---|
| COMPLETE | 4340 (14%) | 560 (13%) | 2580 (14%) | 7660 (13%) | 15140 (14%) |
| ONLY BEFORE | 5520 (18%) | 380 (9%) | 1360 (7%) | 1980 (3%) | 9240 (8%) |
| ONLY AFTER | 4100 (14%) | 1280 (31%) | 7460 (41%) | 20220 (36%) | 33060 (30%) |
| NO USE | 16060 (54%) | 1940 (47%) | 6840 (38%) | 27180 (48%) | 52020 (48%) |
| TOTAL | 30020 (100%) | 4160 (100%) | 18250 (100%) | 57040 (100%) | 109460 (100%) |

(b)

**Figure S6. Hand-hygiene compliance of HCWs and type of room isolation.** (a) Number and (b) cumulative duration of visits to patients in the whole period of data collection, divided according to the use of hydro-alcoholic solution before and after the visit and the type of room isolation.

21